\documentclass[12pt]{article}
\usepackage{psfig}
\begin{document}
\begin{center}
{\Large\bf Generic growth instabilities in one-layered tissue sheets}\\

\vspace{0.5cm}

Dirk Drasdo\footnote{drasdo@imise.uni-leipzig.de} \\
Inst. f\"ur Medizinische Informatik, Statistik und Epidemiologie\\
  Universit\"at Leipzig, Liebigstr. 27, 04103 Leipzig, Germany \\
  and \\
Max-Planck-Institut f\"ur Kolloid-und Grenzfl\"achenforschung, \\
Theory Division, D-14424 Potsdam, Germany
\end{center}

\vspace{1cm}

\begin{abstract}
Growth and folding in one-layered model tissue sheets are studied in a 
stochastic, lattice-free single cell model which considers the discrete 
cellular structure of the tissue, and a coarse grained analytical approach.
The polarity of the tissue sheet is considered by a bending term.
Cell division gives rise to a locally increasing metric.
An exponential and a power-law growth regime are identified.
In both regimes folding occurs as soon as the bending contribution becomes too
small to compensate the destabilizing effect of the cell proliferation.
The potential biological relevance is discussed.
\end{abstract}
Many growth models have been inspired by biological systems or can be applied
to them, see e.g. [1-5].
Most models consider either surface or bulk properties of more or less compact
assemblies of particles (here: cells) as e.g. in Eden-like models of tumor-\cite{Eden},
bacterial-\cite{BenJacob} or yeast growth \cite{Sam}.
We here present a novel microscopic growth model where
cell division can take place only within a single cell layer such, that
a one-layered structure is maintained.
In computer simulations with this model and in a coarse grained continuum 
approach it is shown that the growing layer cannot remain smooth but must fold by a 
mechanism that is believed to be generic, i.e. inherent to this structure.\\
Such situations can occur in all biological cell systems in which cell division
has to maintain a single-layered structure, e.g. in the skin \cite{Farrell},\cite{Pottenunpubl} or in crypts, the 
cell proliferating units in the intestine.
Strong perturbations of the physiological cell division in crypts result
in strong folding \cite{Pottenunpubl},\cite{CairnieMillen1975}, 
producing patterns analogous to those observed during the
formation of polyps or adenoma which constitute pre-patterns of intestinal cancer
\cite{Araki1995}. \\
For simplicity and in order to obtain a clear illustration of the underlying
folding principle we focus on simple
one-dimensional ($1d$) tissue manifolds (''cell chains'') in two-dimensional 
($2d$) space.
Our basic model assumptions are the existence of attractive nearest-neighbor
(NN)
interactions between cells to maintain the integrity of the (one-layered)
tissue sheet, a bending energy that models the polarity of the cells in a sheet
and cell division that takes into account potential size changes of the sheet.
In order to maintain a one-cell thick structure, cells have to be placed in the
sheet again after cell division.
Since cells grow and divide only if this does not result in too strong 
cell deformations or compressions, cell division at constant rate requires the 
migration of either the cell layer itself or of cells within the cell layer.
The growth law and the geometry of the tissue sheet is then determined by the
competition between the area proliferating and destabilizing cell growth and the
stabilizing bending energy of the tissue that locally confines cell movements 
perpendicular to the layer and thus can be responsible for a confinement of cell
division.
As soon as the bending energy becomes too small to smoothen local undulations 
that are stochastically created by local fluctuations in the proliferation of 
length, the layer roughens.
If this occurs before cell deformations or compressions become strong enough
to hinder cell division, the cells divide exponentially fast. 
Otherwise first the layer stays smooth but the growth law changes to 
sub-exponential growth.
Later again, the layer folds.
A large growth rate results in early crossovers.\\
Below we first introduce the microscopic model that shows the crossover between
both growth regimes as well as the folding.
The instability within each growth regime is explained in an analytical 
approach afterwards.
\begin{figure}
\begin{center}
\hspace*{-0.5cm}\psfig{figure=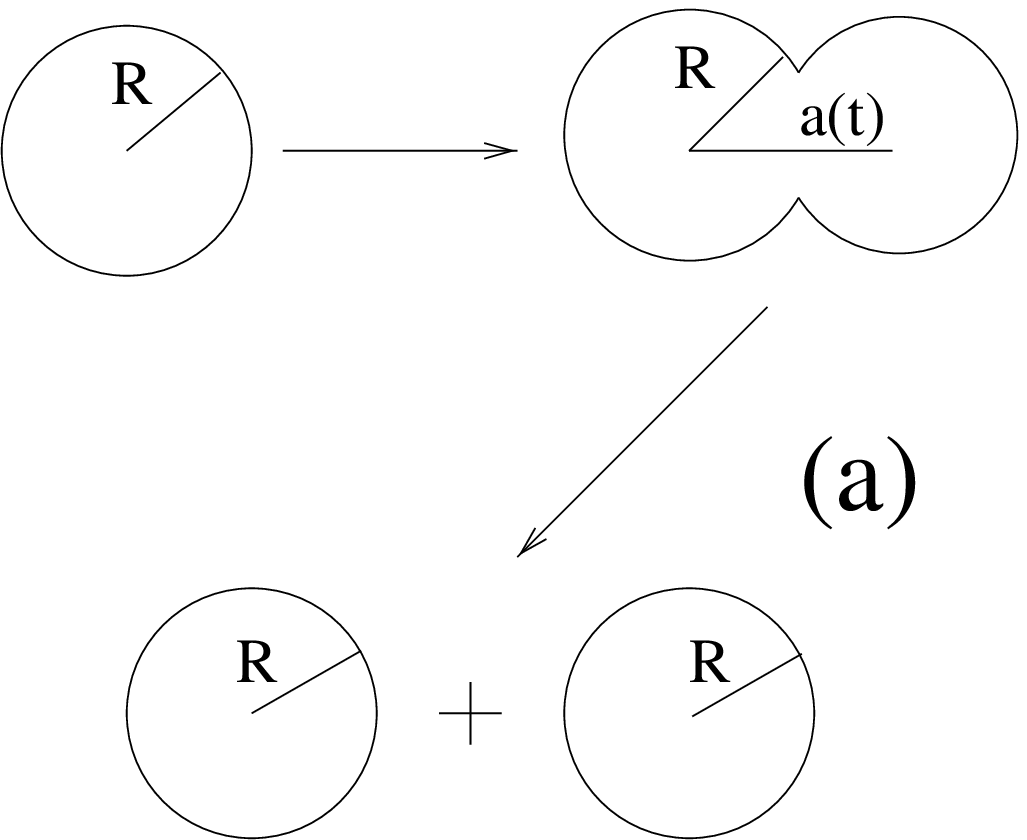,width=5.cm,angle=0}
\hspace*{0.7cm}\psfig{figure=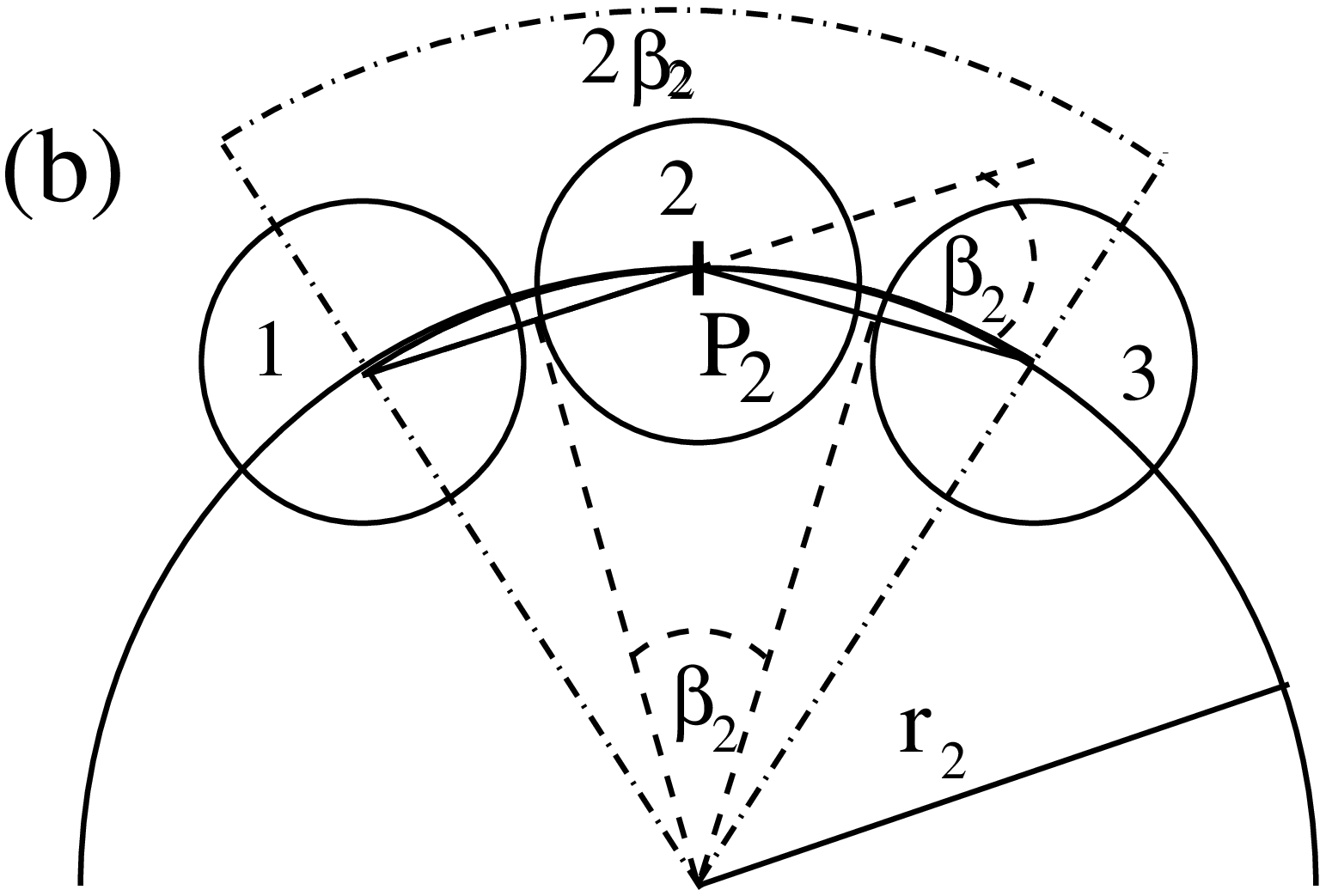,width=5.cm,angle=0}
\end{center}

\noindent
{\small FIG. 1:
(a) Cell division algorithm: 
During growth a cell deforms from a spherical shape into a dumb-bell
by elongation of the axis from $a = 0 \rightarrow a = 2 R$
by small steps.
(b) $r_2$ is the local radius of curvature for cell $i=2$, $\beta_2$ its
curvature angle.
During deformation cell $2$ attempts to orientate its axis into the direction 
of the local tangent to the circle in point $P_2$.
}
\vspace*{-0.3cm}
\end{figure}
The basic unit in the microscopic model is a single cell.
Each cell is assumed to be spherical directly after cell division and deforms 
during mitosis into a dumb-bell (Fig. 1a, \cite{FootnoteGrowth}), i.e. actively changes its 
'equilibrium shape'.
A particular configuration of cells is characterized by its total energy,
$V^{tot} = \sum_{i<j} V_{ij}^{NN} + \sum_{i} V_i^{bend} + \sum_{i} V_i^{rot}$.
The terms on the rhs. characterizes (i) the NN-interaction, 
(ii) the bending contribution, and (iii) non-spheric cells.
We assume that the cell-cell-NN, short-range potential has one minimum
which results from the competition between (a) attractive interactions due to 
adhesion molecules anchored in the cell membranes and (b) repulsive 
contributions from the limited cell deformability and compressibility and the 
loss of membrane steric entropy.
The NN-potential used in the simulations is 
$V_{ij}^{NN} = \epsilon \left((((2d_{ij}(t)-4R)/\delta) - 1)^2 - 1\right)$
for $2R \leq d_{ij}(t) \leq 2R+\delta$ and $V_{ij}^{NN} = \infty$ for 
$d_{ij}(t) < 2R$ or $d_{ij}(t)>2R+\delta$ \cite{FootnoteInfty}.
Here, $d_{ij}(t)$ denotes the distance between the nearest circles of the
neighboring dumb-bells $i$ and $j$ (a circular cell is a dumb-bell with axis 
length $a=0$).
The lower cutoff at $d_{ij}=2R$ takes into account the limited cell 
deformability and corresponds to an excluded volume.
$\delta$ determines the range over which a cell can be stretched or 
compressed in a certain direction.
A typical value is $\delta\approx 0.2R$.
$\epsilon$ ($>0$) is a measure for the resistance against deformations 
($\epsilon\equiv 10$ throughout this article).
We introduce polarity by assuming that an anisotropic distribution of 
cell-cell adhesion molecules gives rise to a bending energy in addition to the 
isotropic energy contribution:
$V_i^{bend} = (1/3)\sum_{j=i-1}^{i+1}(\kappa/2)((1/r_i) - c_0)^2 r_i \beta_i$.
$\kappa$ is the bending rigidity, $c_0$ the spontaneous curvature,
$r_i$ the local radius of curvature and $\beta_i$ the local angle of curvature 
(Fig. 1b).
Cell division in many tissue sheets is directed, e.g. during early 
embryogenesis, in the crypt and the skin, suggesting that cells are 
able to detect the position of their neighbors in order to determine the
direction of their division \cite{WhiteBorisy1983}.
Accordingly we assume that non-spherical cells orientate their axis in a 
way that guarantees the cell division takes place in the direction of the 
tangent to the local radius of curvature (Fig. 1b).
This insures the maintenance of a one-cell thick tissue sheet.
Formally this is carried out by an energy contribution 
$V_i^{rot} = \gamma (\alpha_i - \alpha_i^{opt})^2$ that for $\gamma\gg 1$
sufficiently penalizes deviations from the optimal orientation of the cell axis.
Here, $\alpha_i$ is the momentary, $\alpha_i^{opt}$ the optimal angle of the 
axis in an external coordinate system.\\
Active cell deformations during mitosis cause a pressure on the neighbor cells
in the direction of the deformation.
This leads to an increase of the total energy $V^{tot}$.
We assume that the neighbor cells either move their center of mass 
or change their orientation in order to minimize $V^{tot}$.
Under the assumptions that (i) inertial terms are small compared to dissipative 
terms, (ii) processes not explicitly considered, 
such as the cell metabolism, intra-cellular movements of the cytoplasm and the
reorganization of the cytoskeleton give rise to a stochastic component 
in the displacement of the cells, the dynamics can be modeled by the Metropolis 
method which corresponds to the numerical integration of a Master equation 
\cite{DKM1995}, \cite{Metropolis1953}:
Cells are randomly chosen to perform either a small translation step, 
orientation change or growth trial.
Each translation or rotation is accepted with probability $P_a = 1$ if 
$\Delta V^{tot} = V^{tot}_{t+\Delta t} - V^{tot}_{t} < 0$ and with
probability $P_a = \exp(-\Delta V^{tot}/F_T)$ if $\Delta V^{tot}\geq 0$ 
(hence isolated cells move diffusively in accordance with 
ref. \cite{MombachGlazier}).
$\Delta t$ is the time between two migration or rotation trials ($\equiv 1$ in the
simulations).
$F_T$ is the $k_BT$-analog in cellular systems 
($T$: temperature, $k_B$:Boltzmann const.) \cite{Beysens}
and can be absorbed into a redefinition of $\epsilon$, $\kappa$ and $\gamma$
by $\epsilon\equiv \epsilon/F_T,...\;$.
A growth trial is performed only if it does not result in too strong cell
deformations, i.e. if excluded-volume interactions are irrelevant 
(i.e., if $d_{ij}>2R$).
Between two deformation trials a cell performs $r_g^{-1}$ translation
and rotation trials.
A cell not subject to excluded volume interactions needs $2R/\overline{\delta a}$ growth
steps of average size $\overline{\delta a}$ to once traverse the cell cycle.
Hence its average cycle time is $\tau=(2R\Delta t)/(\overline{\delta a} r_g)$.
Due to rejected growth trials the real average cell cycle time is $\tau_R>\tau$ if 
excluded-volume interactions occur.
In each case a large $r_g$ corresponds to a large cell growth rate and vice 
versa.\\
In $2d$-space one may distinguish between two idealized, extremal 
situations, a cell configuration with fixed ends at $x_1, x_2$ 
(Fig. 2a) and a closed arrangement (Fig. 2b,c).
\begin{figure}\thicklines
\vspace*{0.3cm}
\setlength{\unitlength}{1cm}
\begin{center}
\begin{picture}(15.5,3.5)
\put(-0.4,1.9){\psfig{figure=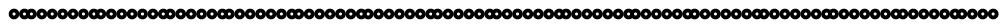,width=3.5cm,angle=0}}
\put(-0.2,1.9){\makebox(0.25,0.25){\large (a.1)}}
\put(-0.3,1.1){\psfig{figure=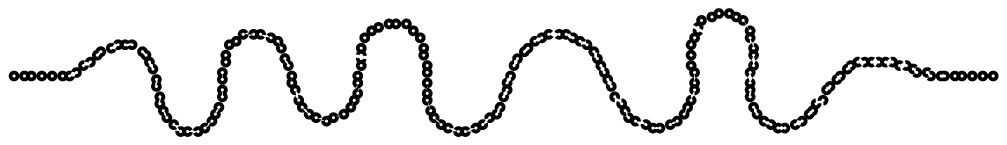,width=3.5cm,angle=0}}
\put(-0.2,0.95){\makebox(0.25,0.25){\large (a.2)}}
\put(1.4,2.2){\vector(0,-1){0.6}}
\put(3.1,2.1){\line(1,-3){0.2}}
\put(3.3,1.5){\vector(-1,-3){0.2}}
\put(-0.3,0.2){\psfig{figure=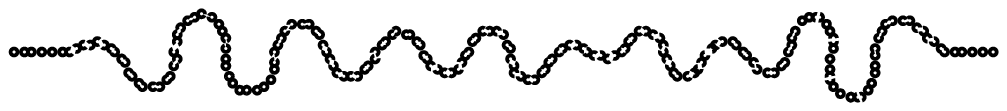,width=3.5cm,angle=0}}
\put(-0.2,0){\makebox(0.25,0.25){\large (a.3)}}
\put(11.6,0.6){\psfig{figure=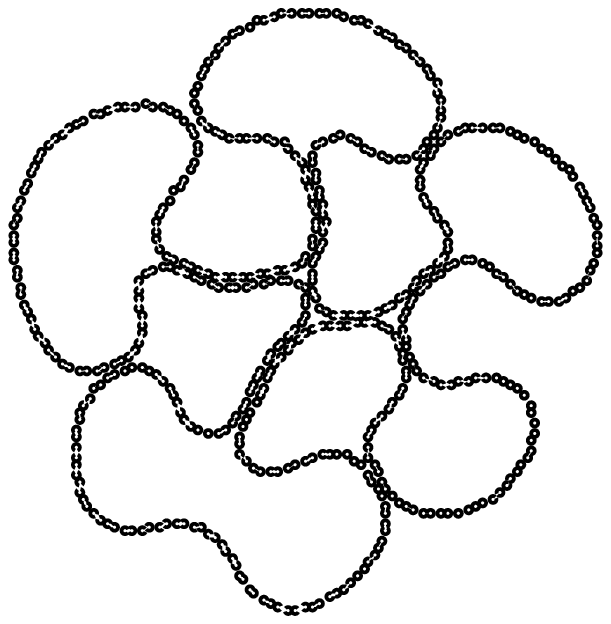,width=2.5cm,angle=0}}
\put(12.6,0.3){\makebox(0.25,0.25){\large (c.1)}}
\put(13.3,0.7){\psfig{figure=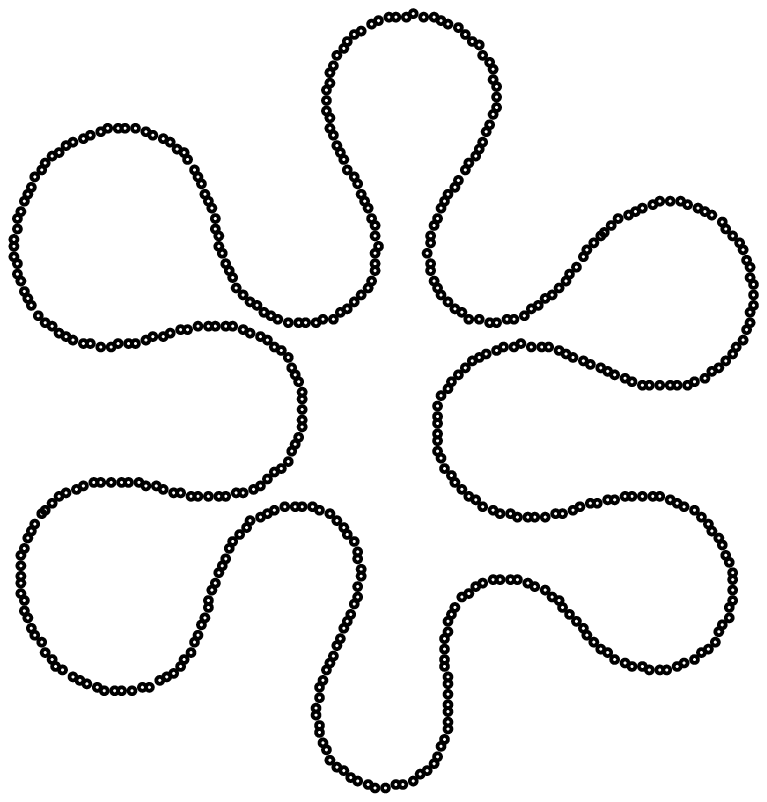,width=2.5cm,angle=0}}
\put(14.5,0.3){\makebox(0.25,0.25){\large (c.2)}}
\put(2.4,0.2){\psfig{figure=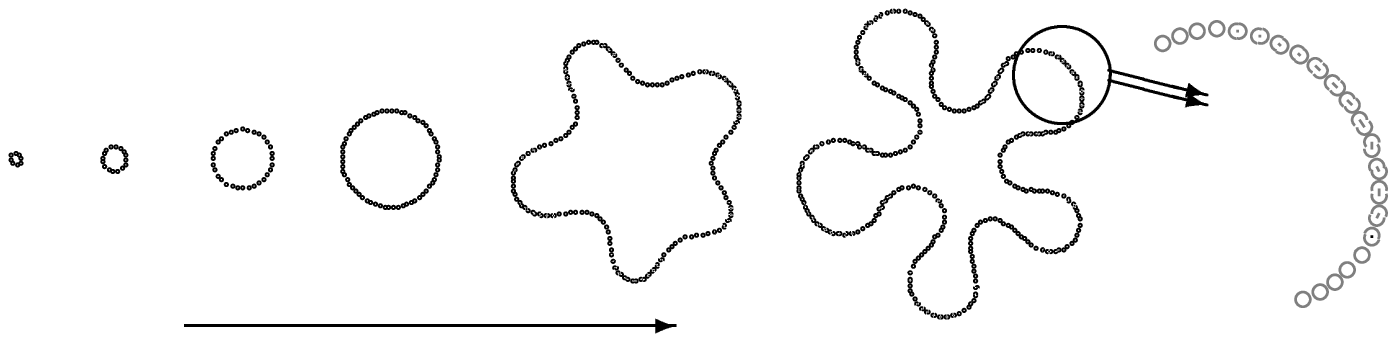,width=10cm}}
\put(4.,1.9){\makebox(0.25,0.25){\large (b)}}
\end{picture}
\end{center}

\noindent
{\small FIG. 2:
(a) Instability in a stretched cell configuration with fixed ends (snapshots,
single arrows indicate the time direction),
(a.1) start configuration, $N=100$, $\kappa=100$ (a.2): 
$r_g^{-1}=200$, $N=110$, (a.3): $r_g^{-1}=50$, $N=110$.
(b) Instability in a circular growing cell arrangement, time series for
$r_g^{-1}=1.25$, $\kappa=2000$ for $N=8,16,32,64,128,256$ (cf. Fig.3b).
(c): snapshots for $N=512$, $r_g^{-1}=500$, 
(c.1): $\kappa=20$, (c.2): $\kappa=100$ (cf. Fig.3a, points A and B).
}
\vspace*{-0.3cm}
\end{figure}
As $U > x_2 - x_1$ ($U$: arc-length of the cell chain), the stretched
configuration becomes curved because the minimum free energy configuration 
as well as the non-equilibrium configurations (Fig.2a) become curved.
As examples of curved configurations we focus in the following on closed
and hence intrinsically curved cell geometries (Fig. 2b,c).
The spontaneous curvature $c_0$ can be dropped in $1d$-closed geometries 
(\cite{Seifert1991} and below).
In each simulation we start with a circular configuration of eight cells.
For small times $t$ the circumference $U$ 
and the total number of cells $N$ grow exponentially fast 
$U \propto N \propto \exp(\lambda t)$ with $\lambda=(\tau ln(2))^{-1}$
(Fig.3a) \cite{FootnoteBlast}.
Due to the excluded volume effect exponential cell division is 
accompanied by exponentially fast depletion of free space on time scales 
$t > \tau$.
Hence to maintain exponential growth at still circular geometry the migration 
of each cell $i$ into the direction of the locally outward pointing normal 
$\underline{n}_i$ $\forall i=1,..,N$ would have to be exponentially fast, too.
The driving force $\underline{F}_i$ for the migration is mainly given by the 
repulsive part of
$V_{ij}^{NN}$ since due to the elongation of the cell axis and the introduction
of new cells, the cell-cell-distance is reduced to $d_{ij}<2R+\delta/2$.
Its component $F_{n_i}$ into the direction of $\underline{n}_i$ is 
$F_{n_i} \equiv F_n \propto \beta \propto 1/r$ for $r \gg R$ thus decreases 
with $r$ (for circular configurations we can drop the index $i$ here).
\begin{figure}
\setlength{\unitlength}{1cm}
\hspace*{0.6cm}\psfig{figure=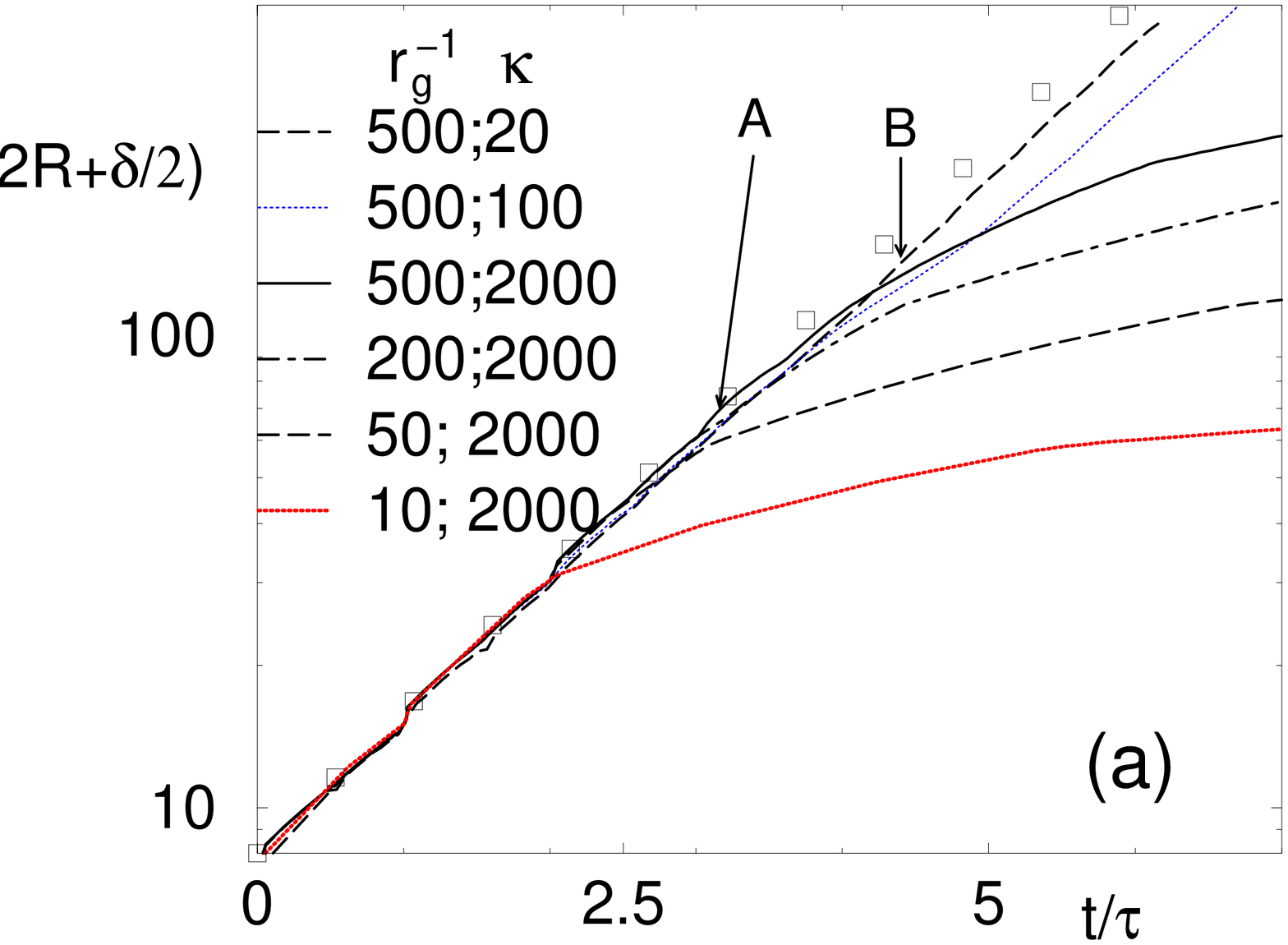,width=6.3cm,angle=270}
\hspace*{1cm}\psfig{figure=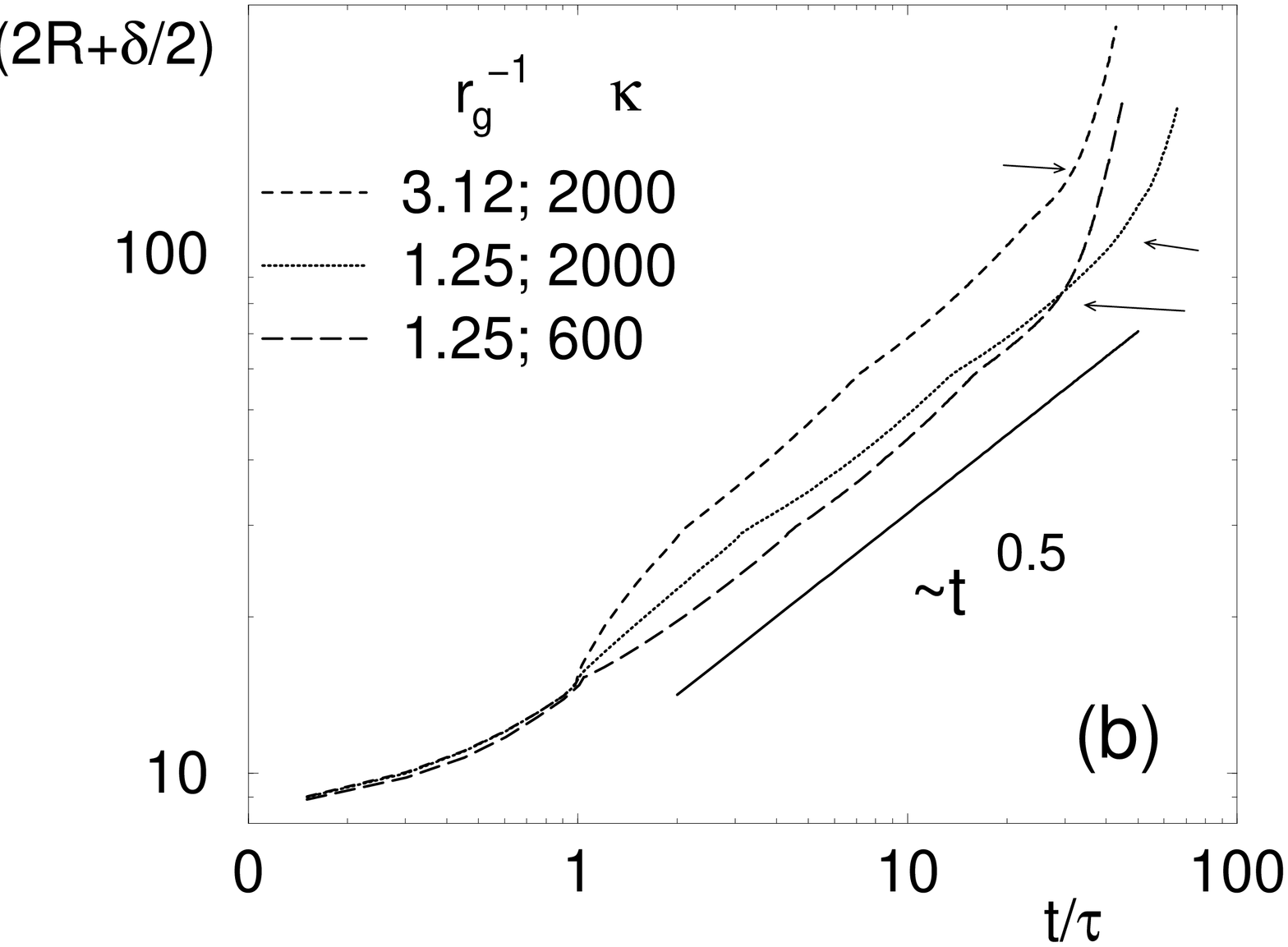,width=6.5cm,angle=270}

\noindent
{\small FIG. 3:
Circumference $U$ in units of $2R+\delta/2$ vers. $t/\tau$ on lin-log (a) and
log-log-scale (b). 
(a) the initial exponential regime is the longer the smaller are $\kappa$ and 
$r_g$ (squares: $\propto 2^{t/\tau}$). 
For small $\kappa$ and $r_g$ a crossover occurs in the 
geometry (e.g. A:$r_g^{-1}=500$, $\kappa=20$, B: $\kappa=100$ in (a)),
for large $\kappa$ first in the growth law changing from an 
exponential to a power law ($\propto t^{\alpha}$ with $\alpha\approx 0.5$;
$\tau_R > \tau$) (b).
For sufficiently large times $t$, $U$ is bent upwards indicating a second
crossover to a non-circular configuration.
The domain size at the crossover is the smaller, the smaller are $r_g^{-1}$ and 
$\kappa$ (cf. arrows in (b)).
}
\end{figure}
Since the migration dynamics is strongly dissipative, the force does not suffice
to relax the cell-cell distance to its equilibrium value between two successive
cell divisions.
For small $\kappa$, the configuration then roughens at still exponential growth
(scenario (1), Fig. 3a).
For large $\kappa$ the cell-cell distance shrinks until it reaches the lower 
cutoff-value $d_{ij}=2R$. 
After this happens only those migration trials which fall into the angle 
interval $\beta \propto 1/r$ are successful, others would result in a 
forbidden large deformation of the cells (i.e., $d_{ij}<2R$).
From $dr/dt \propto F_{n} \propto 1/r$ we expect a change of the growth law to
$U \propto r \propto \sqrt{t}$ at still perfectly circular cell 
configurations.
This is indeed found within some limits from the crossover effects (scenario (2), 
Fig. 3b).
For large times a second crossover occurs at which the cell configuration 
starts to deviate significantly from a perfect circle, indicated as a 
characteristic bend upwards in the growth curve (Fig. 3b).
The instabilities occur at typical domain sizes that grow with 
$r_g^{-1}$ (Fig. 2a; arrows in Fig. 3b) and $\kappa$ (Fig. 2c,
points (A) and (B) in Fig. 3a).\\
The nature of the geometric instability can best be understood in a simplified
continuum model.
Connecting the middle-points of the cells in a given configuration
we arrive at a closed curve in $2d$-space.
As in ref. \cite{GoldsteinLanger1995} we parameterize 
the curve by the position vector
$\underline{r}(\alpha,t)$ where $\alpha \in [0,1)$ is a parameter, $t$ the
time.
$\underline{r}(\alpha,t)$ and its first two derivatives are periodic in 
$\alpha$.
Note that $U = \int_0^1 \sqrt{g} d\alpha$ where
$g =\partial_{\alpha} \underline{r}(\alpha,t)
\partial_{\alpha} \underline{r}(\alpha,t)$ is the determinant
of the metric tensor.
The mechanisms that contribute to the dynamics are 
(a) local proliferation of arc-length due to cell divisions,
(b) stabilization of stretched structures by a bending energy, (c) constant
(at least on the average) circumference if cell division is switched off.
We only consider cell configurations near the instability where
intersections of the curve with itself cannot occur. Then,
\begin{eqnarray}
\frac{1}{\mu}\left. \frac{\partial \underline{r}(\alpha,t)}{\partial t}\right|_{\alpha} =
-\frac{1}{\sqrt{g}} \frac{\delta {\cal F}}{\delta \underline{r}}.
\label{DEDyn}
\end{eqnarray}
$\mu$ is the mobility. The prefactor $1/\sqrt{g}$ ensures re\-para\-metri\-za\-tion 
invariance.
${\cal F} = {\cal F}_0(\alpha,t) - \int_0^1\Lambda(\alpha,t)\sqrt{g}d\alpha$
where $\Lambda(\alpha,t)$ is a Lagrangian multiplier field which ensures that
the condition for the local proliferation of length, that 
is specified below, is fulfilled.  
${\cal F}_0 =\kappa/2 \int_0^1 (c(\alpha,t) - c_0)^2 \sqrt{g} d\alpha$
is the bending energy ($c(\alpha,t)$ is the local, $c_0$ the spontaneous
curvature).
To get a closed set of equations,
we now have to fix a condition for the local metric by 
$\partial_t\sqrt{g} =\partial_{\alpha}\underline{r}\partial_{\alpha}\partial_t\underline{r}/\sqrt{g} \equiv f$.
$f$ describes the growth of the metric and has to be given externally.
With the Serret-Frenet relations \cite{GoldsteinLanger1995} we find
$\mu \sqrt{g}
(c W_n + \partial_s W_t + c^2 \Lambda - \partial^2_s \Lambda) = f$
with $W_n= -(1/\sqrt{g})(\delta{\cal F}_0/\delta\underline{r})\underline{n}$
$=(\kappa/2)(c^3+2\partial_s^2c-cc_0^2)$ (if $\partial_{\alpha}c_0=0$) and 
$W_t=-(1/\sqrt{g})(\delta{\cal F}_0/\delta\underline{r})\underline{t}=0$.
$f$ has been chosen to 
$f\equiv f_{EGR}=\sigma\sqrt{g}$ in the exponential growth regime (EGR) 
and $f\equiv f_{PGR}=\sigma\sqrt{g}c^2$ in the power-law growth regime 
(PGR).
$\sigma$ measures the strength of growth.
The choice of $f_{EGR}$ insures exponential growth of the metric, 
the choice of $f_{PGR}$ can be shown to correspond to a  
normal growth velocity $G_n\propto c$ ($\propto \beta$ for $r\gg R$, cf. 
Fig. 1b) and to a transversal velocity $G_t$ that obeys the relation $\partial_s G_t=0$
(from a transformation $\lambda=\Lambda + G_n/c$ and $G_t\equiv-\partial_s(G_n/c)$).
Note that $c_0$ can be absorbed into a redefinition of 
$\Lambda \equiv \Lambda-(\kappa/2)c_0^2$.
These settings result in the homogeneous eqns. $\partial_t r = \sigma r\Rightarrow r\propto e^{\sigma t}$ for EGR and 
$\partial_t r = \sigma/r\Rightarrow r\propto\sqrt{t}$ for PGR in 
accordance with the computer simulations.
Performing a linear stability analysis around a homogeneous growing circle
(i.e., in polar coordinates: $r(\alpha,t)=r_0(t)+\xi_0\exp(w(q)t+i 2\pi q\alpha)$,
$\Lambda(\alpha,t)=\Lambda_0(t)+\eta_0\exp(w(q)t+i 2\pi q\alpha)$ 
where $r$ determines the dynamics of $\Lambda$), we find
\begin{eqnarray}
w(q) =
\left\{
\begin{array}{ccc}
-\frac{\mu\kappa q^2(q^2 - 1)^2}{r_0^4(1 + q^2)} + 
\frac{\sigma(q^2 - 1)^2}{1 + q^2} & for & EGR \\
-\frac{\mu\kappa q^2(q^2 - 1)^2}{r_0^4(1 + q^2)} + 
\frac{\sigma(q^2 - 1)}{r_0^2} & for & PGR \\
\end{array}
\right.
\label{Approxw}
\end{eqnarray}
$w(q=0)=\sigma>0$ for EGR, so homogeneous deviations grow. 
For PGR $w(0)=-\sigma/r_0^2$, so homogeneous perturbations are damped out.
$w(q=1)=0$ which implies that a translation of the circle is marginally stable
\cite{Seifert1991}.
A second zero is at $q_c=\sqrt{X(2)}$ for EGR and
$q_c=\sqrt{(1+X(1)+\sqrt{1+6X(1)+X(1)^2})/2}$
for PGR where $X(z)=r_0^{2z}\sigma/(\mu\kappa)$.
For $q\rightarrow\infty$, $w(q)<0$ i.e., short wavelength perturbations are 
damped out.
For $\sigma=0$ the circle is the only stable solution.
The fastest growing mode is $q_m \approx \sqrt{X(z)/2}$ for $q_m\gg 1$.
$q_c$ and $q_m$ grow with increasing $X(z)$, i.e. the domain sizes at the
geometrical instabilities decrease with increasing growth rate and decreasing 
bending rigidity.
These results are in agreement with the tendencies found in the computer
simulations.
From the analytical treatment it becomes clear that the instability occurs at
domain sizes where the bending energy does not suffice to smooth the
roughening effect of the cell proliferation. 
An increase of the proliferation rate has the same effect: the ''time''
between two subsequent cell division doesn't suffice anymore to
smoothen local undulations.
For PGR, cells at positions with larger than average local curvature have a smaller
than average mitotic cycle thus divide faster than cells at positions with 
smaller local curvatures.
This further increases the difference in the local curvatures in a 
self-enhancing process resulting in a folding of the domain.
For EGR, the increment in the metric is proportional to the metric already
proliferated.
The folded structures are not in equilibrium:
for the $1d$-ring structures, the undulations regress if the cell
division is suppressed.
There is no contribution due to shear. \\
Also $2d$-buds formed in $3d$ may reorganize into a perfectly smooth layer on large 
time scales and reduce shear energy. 
This line of argument is supported by observations which show that
cell assemblies can behave as viscoelastic fluids: under compression between
two plates, cell assemblies first deform followed by a reorganization of the 
cells in order to reduce elastic energy contributions \cite{ForgacsBPJ}.
A separate experiment identified cells to diffuse even in aggregates \cite{MombachGlazier}.
However, growth in a $2d$-surface for some cases is expected to be
at least temporarily accompanied by shear stress that elevates the energy barrier 
for the instability, but is not expected to change its nature.\\
\hspace*{0.25cm} The folding principle neither depends on details of the 
cell division algorithm nor on the particular shape of the NN-potential 
suggesting some universal feature \cite{DrasdoInPrep}.
Hence this mechanism may be present in all one layered epithelial tissues
including those embedded in soft connective tissue during development and the
maintenance of tissue if the rate of proliferation is sufficiently large either 
by internal or external stimuli.
A typical biological example for the folding mechanism may be the 
post-irradational situation in crypts \cite{Pottenunpubl},\cite{CairnieMillen1975}.
Crypts form pear-shaped pockets in the intestinal wall 
\cite{PottenLoeffler1990}.
They are responsible for the maintenance of the high cell turnover in the
intestine.
After irradiation the cell proliferation increases and a fingering-like 
instability occurs in their floor or walls.
Folding is also observed in different geometries such as the skin after 
treatment with growth factors or during the molting and growth of the epidermis 
of arthropods under hormonal control \cite{Wolpert}. \\
For a particular biological situation all simulation parameters can 
be related to experimental quantities \cite{DrasdoForgacs}. \\
DD. is very grateful to M. Loeffler for stimulating discussions and 
bringing refs. \cite{CairnieMillen1975}, \cite{Araki1995} into his attention.
He further thanks G. Forgacs, M. Kschischo, C.S. Potten, U. Seifert and 
especially G. Gompper for discussions or suggestions to the manuscript.


\begin{thebibliography}{21}
\bibitem{FootnoteInfty}
Reversible cell-cell contacts can be modeled by $V_{ij}^{NN} = 0$ if 
$d_{ij}>\delta$.
Closed cell geometries (cf. Fig. 2bc,3) are then broken up due to $V^{bend}$
if $\epsilon \leq 2\pi^2\kappa/((2R+\delta/2)N)$.
\end{thebibliography}
\end{document}